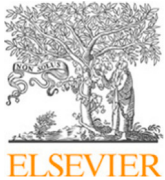
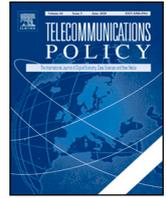

# Framing metaverse identity: A multidimensional framework for governing digital selves

Liang Yang [a], Yan Xu [b,*], Pan Hui [c]

[a] *Division of Emerging Interdisciplinary Areas, The Hong Kong University of Science and Technology, Hong Kong Special Administrative Region of China*
[b] *Department of Information Systems, Business Statistics and Operations Management, The Hong Kong University of Science and Technology, Hong Kong Special Administrative Region of China*
[c] *Computational Media and Arts Thrust, The Hong Kong University of Science and Technology (Guangzhou), Guangzhou, China*

A B S T R A C T

This paper proposes a multidimensional framework for *Metaverse Identity*, addressing its definition, guiding principles, and critical challenges. *Metaverse Identity* is conceptualized as a user's digital self, encompassing personal attributes, data footprints, social roles, and economic elements. To elucidate its core characteristics and implications, this framework introduces two guiding principles: *Equivalence and Alignment* and *Fusion and Expansiveness*. The first principle advocates for consistency between metaverse and real-world identities in behavioral norms and social standards, ensuring rights protection and establishing conduct guidelines. The second emphasizes the deep integration and transformative evolution of metaverse identities, enabling them to transcend real-world constraints, meet diverse needs, and foster inclusivity. Together, two principles serve as complementary pillars, balancing ethical integration with dynamic co-evolution. Building on this foundation, the study identifies five critical challenges: interoperability, legal boundaries, privacy and identity management, risks from deepfakes and synthetic identities, and identity fragmentation impacting psychological well-being. To address these challenges, strategic recommendations are offered to guide stakeholders. By constructing this framework, the study fills a key theoretical gap, advances systematic research, and provides a foundation for policies and governance strategies to address the complexities of metaverse identities in a rapidly evolving digital domain.

## 1. Introduction

The metaverse is envisioned as the future of the Internet, aspiring to offer a spatial and social Internet experience that seamlessly blends the physical and digital realms using both existing and emerging technologies (Hollensen, Kotler, & Opresnik, 2022; Koohang et al., 2023; Venugopal, Subramanian, & Peatchimuthu, 2023). While the term has gained prominence in recent years, largely due to its promotion by industry leaders such as Meta (formerly Facebook) (Peukert et al., 2024), it has also garnered significant academic interest. Particularly in the last three years, scholarly research on the metaverse has extensively covered various dimensions, including but not limited to technological advancements (Fu et al., 2023; Huynh-The et al., 2022; Park & Kim, 2022b; Xu et al., 2022), ecosystem dynamics (Lee et al., 2021), social impacts (Dwivedi et al., 2022; Hennig-Thurau, Aliman, Herting, Cziehso, & Linder, 2022; Park & Kim, 2022b; Šímová, Zychová, & Fejfarová, 2023), economic potential (Bhattacharya et al., 2023), and user

* Corresponding author.
*E-mail addresses:* lyangbl@connect.ust.hk (L. Yang), xuyan@ust.hk (Y. Xu), panhui@ust.hk (P. Hui).

behavior and experiences (Buhalis, Lin, & Leung, 2022; Park & Kim, 2022a; Riva & Wiederhold, 2022).

The term *metaverse* was coined by Neal Stephenson in his 1992 novel *Snow Crash*, where he conceptualized it as a virtual reality application. In his depicted *STREET* metaverse, humans as avatars interact with intelligent agents and each other in an immersive virtual world (Dionisio, Iii, & Gilbert, 2013; Joshua, 2017). Mark Zuckerberg, the founder of Meta, described the metaverse as an integrated immersive ecosystem; the virtual and real worlds merge seamlessly, enabling users to engage in shared experiences through avatars and holograms in this vision (Dwivedi et al., 2022). However, there is no consensus in academic literature on the definition of metaverse (Almoqbel, Naderi, Wohn, & Goyal, 2022; Dwivedi et al., 2022; Park & Kim, 2022b; Peukert et al., 2024; Ritterbusch & Teichmann, 2023). Schultze views it as a dynamic concept emerging from the convergence of experiential computing practices and argues for research focused on practical implications and user experience impacts (Peukert et al., 2024). While Schultze emphasizes the metaverse's evolving nature, this study adopts Matthew Ball's definition for its clarity and comprehensiveness. Ball describes the metaverse as "a massively scaled and interoperable network of real-time rendered 3D virtual worlds that can be experienced synchronously and persistently by an effectively unlimited number of users with an individual sense of presence, and with continuity of data, such as identity, history, entitlements, objects, communications, and payments" (Ball, 2022). This definition conceptualizes the metaverse as a vast network of virtual worlds, heralding a new era of convergence between virtual and physical realities.

Compared to earlier eras of the Internet and virtual worlds, the metaverse represents a new frontier in empowering self-expression (Peukert et al., 2024; Roblox, 2023). The evolution of self-expression has closely tracked technological advancements, illustrating a dynamic interplay between innovation and identity formation. From usernames in the early Internet era (Huffaker & Calvert, 2005) to curated social media profiles (Gardner & Davis, 2013; Lewis & Fabos, 2005; Zhao, Grasmuck, & Martin, 2008) and avatar-based representations in virtual worlds (Bullingham & Vasconcelos, 2013; Schultze, 2010; Schultze & Leahy, 2009), digital technologies have not only enabled individuals to explore and express their identities but have also reshaped how identities are constructed and perceived (Adhiarso, Utari, & Hastjarjo, 2019; Hoehe & Thibaut, 2020; Romeo, 2024; Thorne, Sauro, & Smith, 2015). The metaverse elevates self-expression to unprecedented levels by offering immersive, authentic, and multifaceted digital experiences that simulate real-world interactions (Lee, Yang, & Hui, 2023; Mystakidis, 2022; Yang, Ni et al., 2024). Within these spaces, individuals can embody multidimensional aspects of their identities (Lee, Yang, & Hui, 2023; Peukert et al., 2024; Roblox, 2023), transforming digital identity from an abstract concept into a tangible, embodied experience akin to physical reality (Ball, 2020). However, this empowerment comes with challenges. Identities in the metaverse, like those in the physical world, face a dual nature: they provide opportunities for free self-expression and creativity but also amplify concerns around data privacy, cybersecurity, and systemic biases, which can significantly impact user trust and well-being (Dwivedi et al., 2022; Lake, 2019; Mitrushchenkova, 2023; Schulenberg, Freeman, Li, & Barwulor, 2023).

The challenges associated with *Metaverse Identity* have become a critical focus in academic and policy discussions. The International Criminal Police Organization (INTERPOL) warns that the construction and governance of identity within the metaverse have already emerged as pressing policy concerns, citing the rise of "meta-crimes",—new forms of criminal activity that underscore the urgent technological, social, and ethical complexities of this evolving domain (INTERPOL, 2024). Similarly, the World Economic Forum (WEF) cautions that failing to address the evolution of metaverse identity risks perpetuating existing flaws of the Internet era, such as privacy violations, security vulnerabilities, and unequal access to opportunities, while also exacerbating psychological and emotional risks for users (WEF, 2024). Identity has also been identified as a cornerstone of metaverse interoperability. Yang, Ni et al. (2024)'s systematic review emphasizes that current approaches to metaverse identity remain underdeveloped and require further research to address these gaps. Likewise, Peukert et al. (2024) stress the need for new concepts and frameworks, particularly those focusing on the formation, management, and perception of digital identities within the metaverse. Despite these efforts, a significant research gap persists in developing a multidimensional framework for understanding metaverse identity (also referred to as "meta-physical identity") that incorporates diverse stakeholder perspectives (WEF, 2024). Such a framework is essential for addressing the governance, ethical, and technological challenges posed by the metaverse and ensuring its equitable and sustainable development.

This study conducts an exploration of the key existing literature on identity, digital identity, and metaverse identity to establish a guiding framework for diverse stakeholders to comprehend the distinct features and wider impacts of *Metaverse Identity*. Based on the theoretical foundation, it defines *Metaverse Identity* as the multidimensional construct of a user's digital self (Turkle, 1995, 1999), integrating personal attributes (Lee et al., 2021; Lee, Chang, Uhm, & Owiro, 2023; Lee, Yang, & Hui, 2023; Park & Kim, 2022b; Saker & Frith, 2022; Schultze, 2010; Schultze & Leahy, 2009; Schultze & Orlikowski, 2010; Spence, 2008; Šímová et al., 2023; WEF, 2024), data footprints (Falchuk, Loeb, & Neff, 2018; Tao, Dai, Xie, & Wang, 2023; WEF, 2024; Yang, Zhang, Tian, & Ma, 2023; Zichichi et al., 2023), social roles (Calongne, Sheehy, & Stricker, 2013; Hollensen et al., 2022; Schultze, 2014, 2016; Šímová et al., 2023; Yang, Kim, Song, Lee, & Jang, 2024), and economic elements (Lee, Yang, & Hui, 2023; Momtaz, 2022; Tao et al., 2023; Truong, Le, & Niyato, 2023; Wang et al., 2024; Yang et al., 2022) within the metaverse. This construct reflects the intricate interplay between an individual's online and real-world identities, encompassing all aspects of their digital selves, such as self-presence (Lee, Yang, & Hui, 2023) and self-expression (Peukert et al., 2024; Roblox, 2023) within the metaverse. To enhance understanding of *Metaverse Identity* and its implications, the study proposes two guiding principles—*Equivalence and Alignment* and *Fusion and Expansiveness*. The principle of *Equivalence and Alignment* emphasizes the need for ethical standards and behavioral norms in the metaverse that correspond to those of the real world, ensuring accountability and the safeguarding of individual rights (Bellini, 2024; Dwivedi et al., 2022; Lake, 2019; Lee, Yang, & Hui, 2023; Mitrushchenkova, 2023). Conversely, the principle of *Fusion and Expansiveness* highlights the dynamic co-evolution of virtual and real-world identities, enabling diverse identity expressions, personal growth, and creative innovation (Al-kfairy, Alomari, Al-Bashayreh, & Tubishat, 2024; Ball, 2020; Hatada et al., 2024;

Mitrushchenkova, 2023; Schultze, 2014, 2016; Yang, Kim, et al., 2024). These two principles function as complementary pillars for understanding *Metaverse Identity*, balancing ethical integration with dynamic co-evolution. This study also identifies five critical challenges requiring immediate attention: identity interoperability (Alizadeh, Andersson, & Schelén, 2022; Jiang, Cao, & Wu, 2023; Lee, Yang, & Hui, 2023; Wang et al., 2022; Yang, Ni et al., 2024), legal and regulatory implications (Bellini, 2024; Cheong, 2022; Lee, Yang, & Hui, 2023; Rigotti & Malgieri, 2024), privacy and identity management (Nair, Guo et al., 2023; Nair, Rack et al., 2023; Nair, Rosenberg, O'Brien, & Song, 2024; Rosenberg, 2023), the proliferation of deepfakes and synthetic identities (Gonçalves, Coelho, Monteiro, Melo, & Bessa, 2022; Qayyum et al., 2024; Tariq, Abuadbba, & Moore, 2023; Wu, Hui, & Zhou, 2023), and the risks of identity fragmentation to psychological well-being (Banerji, 2023; Koohsari et al., 2023; Yang, Kim, et al., 2024). The study outlines general potential strategies to address these challenges, offering actionable insights for stakeholders. Through this analysis, the paper constructs a multidimensional framework to bridge theoretical gaps in existing research on *Metaverse Identity*. This framework provides a foundation for advancing future systematic and empirical studies while providing a basis for formulating policies and governance strategies to tackle the multifaceted challenges of this rapidly evolving digital domain.

The structure of the paper is as follows: Section 1 introduces the research background, emphasizing our framework and its contribution of our work to address *Metaverse Identity*. Section 2 examines definitions of identity, digital identity, and metaverse identity, providing a conceptual foundation that traces the evolution of digital identity from *Internet* to *Virtual Worlds* to *Metaverse*. This section summarizes key literature and defines *Metaverse Identity* as a theoretical basis for subsequent discussions. Section 3 distills two core principles for understanding *Metaverse Identity* and illustrates them using representative cases. This approach is designed to help diverse stakeholders, including those without deep expertise in metaverse technologies and theories, develop an intuitive understanding. It also suggests strategies for integrating these principles while critically examining their limitations. Section 4 identifies five critical challenges related to *Metaverse Identity*, providing guidance for applying the principles in technological innovations, ethical considerations, and policy developments. Section 5 reiterates the study's limitations and main findings, aiming to inspire and guide future research in this rapidly evolving field.

## 2. Theoretical foundations

The investigation of *Identity* represents a crucial philosophical inquiry into human existence. At a fundamental level, *Identity* encompasses personal details about who a person is, such as name, age, gender, place of origin, and nationality. However, the Cambridge Dictionary expands this definition to include it as "the fact of being, or feeling that you are, a particular type of person, organization, etc.; the qualities that make a person, organization, etc. different from others".[1] This broader perspective aligns with philosophical explorations of the 'Self', epitomized by questions like "Who am I? Where do I come from? Where am I going?" (Mitrushchenkova, 2023). Historically, philosophers like Aristotle have recognized that *Identity* is shaped both biologically and socially—humans are not merely biological entities but also subjects within social contexts (Mitrushchenkova, 2023). In contemporary discourse, the concept of *Identity* has gained significance with developments in psychiatry, psychology, and sociology. For example, Turkle (1995) in "Life on the Screen" pioneered the exploration of the relationship between technology and identity by investigating how digital technologies interpret and describe human behavior. Since then, extensive research has demonstrated that digital technologies shape our identities and influence how we interact with them (Adhiarso et al., 2019; Golubeva, 2020; Hoehe & Thibaut, 2020; Rani, 2022; Romeo, 2024; Thorne et al., 2015; Turkle, 1999). Later, Ching and Foley (2012) in "Constructing the Self in a Digital World" argues that *Identity* is a dynamic framework and a continuously evolving, fluid process within digital contexts.

Subsequently, numerous studies have explored how digital technologies and experiences shape identity and, conversely, how identity reshapes these technologies. Digital environments such as blogs (Huffaker & Calvert, 2005), social networking sites (Gardner & Davis, 2013; Zhao et al., 2008), instant messaging platforms (Lewis & Fabos, 2005), video-sharing websites (Hu, Zhang, & Luo, 2016; Yang & Wang, 2015), video blogs (Griffith & Papacharissi, 2010), and virtual worlds (Bullingham & Vasconcelos, 2013; Schultze, 2010, 2014, 2016; Schultze & Leahy, 2009) provide us with a multitude of tools for identity exploration and expression. These tools helps us explore, express, and shape our identities. For example, research focusing on social networks like Facebook illustrates how these tools allow users to overcome physical limitations and create identities that would otherwise be unattainable (Gardner & Davis, 2013; Zhao et al., 2008).

Among the digital platforms, virtual worlds such as Second Life, Fortnite, and Roblox are considered precursors to metaverse platforms (Dwivedi et al., 2022; Gent, 2022). Research works by (Schultze, 2010, 2014, 2016) have offered critical insights into identity expression in virtual worlds and their relationships with real-world identities. Through her theoretical perspective of "*Performative Identity*", Schultze elucidates the complexity of identity manifestation in these environments. She posits that users' identity performances in virtual worlds are not mere replications of their physical selves, but are dynamically generated through avatar-mediated interactions. Avatar-driven behaviors, including sartorial choices, gestures, and postures, reflect users' real-world identity practices while facilitating self-reconstruction within virtual contexts. This dynamic identity performance enables users to extend their self-perception and expression through technological mediation, blurring the lines between virtual and physical realms, as well as between technology and human agency (Schultze, 2014).

Schultze's conceptualization of "*Performing Cyborgian Identity*" further illuminates the technological extension of users' sense of self and identity. Identity performance in virtual worlds influences digital interactions and reciprocally shapes identity cognition and behavior in physical reality. For instance, Schultze's case studies in Second Life demonstrate how virtual identities impact users' self-perception, revealing a bidirectional interplay between virtual and real-world identities (Schultze, 2016). This feedback mechanism suggests that virtual worlds function not merely as platforms for identity expression but spaces for self-exploration and development. Environments like Second Life serve as crucial social laboratories, facilitating the exploration of novel roles and competencies while enriching discourse on the interrelation between virtual and physical identities. Schultze's research demonstrates that virtual identity is not simply an extension of physical identity but a dynamic, fluid process of identity construction (Schultze, 2010, 2014, 2016; Schultze & Leahy, 2009; Schultze & Orlikowski, 2010). Through interaction with real-world identities, virtual worlds provide a space for continuous redefinition of self and social roles. This bidirectional interaction offers a theoretical foundation for comprehending identity construction in the emergent metaverse paradigm.

---

[1] https://dictionary.cambridge.org/dictionary/english/identity

While the definition of the Metaverse remains dynamic, the prevailing consensus is that it relies on cutting-edge technologies such as virtual reality (VR), augmented reality (AR), artificial intelligence (AI), and blockchain, key components of metaverse technologies, which radically transform our perceptions, experiences, and interactions with the world (Dwivedi et al., 2022; Lee et al., 2021; Lee, Yang, & Hui, 2023; Mystakidis, 2022; Peukert et al., 2024). As the distinction between digital and real-world identities becomes indistinct, understanding the interaction between digital and real-world identities, as well as the changes brought to them by the metaverse, is paramount. Compared to the *Internet* and *virtual worlds*, the metaverse provides a more immersive, authentic, and diverse digital lifestyle, enabling users to engage with the digital world, closely mimicking real-world experiences (Bradley & Muñoz, 2024; Lee, Yang, & Hui, 2023; Mystakidis, 2022; Peukert et al., 2024; Yang, Ni et al., 2024). For instance, within these virtual environments, users can project multidimensional information – such as preferred physical traits, natural movement habits, and genuine emotional states – through their digital avatars, thereby presenting themselves in a more vivid and three-dimensional manner (Lee, Yang, & Hui, 2023; Peukert et al., 2024; Roblox, 2023). The metaverse transforms identity from a mere concept to a lived experience, highlighting the need for a delicate balance between self-expression and privacy protection (Ball, 2020). In the real world, identity encompasses diverse facets such as race, age, occupation, culture, hobbies, gender identity, and sexual orientation, which are crucial for self-presence, self-expression, and pride. Yet, these same attributes can also expose individuals to risks such as bullying, harassment, stalking, discrimination, litigation, legal actions, persecution, deception, and bias (Dwivedi et al., 2022; Lake, 2019; Mitrushchenkova, 2023; Schulenberg et al., 2023).

*Identity* in the metaverse inherits this duality from the real world, yet it becomes even more complex. The metaverse provides users with a creative space unrestricted by physical rules, allowing them to interact with other individuals and environments across various dimensions (Lee et al., 2021; Qin, Wang, & Hui, 2025; Wider et al., 2023). This freedom enables users to craft unique digital experiences, unleash their creativity, and rethink their goals, desires, and identities (Mitrushchenkova, 2023). However, the complexity of identities in the metaverse emerges not only from their multi-layered nature but also from their dynamic evolution and diversity. Users can choose to present different identities in various virtual environments and continuously adjust and change identity characteristics within the same environment (Mitrushchenkova, 2023). Such fluidity and plasticity offer users unprecedented freedom of self-expression, but they also pose significant privacy and security challenges (Ball, 2020; Dwivedi et al., 2022; Lee et al., 2021). Neglecting identity considerations in the metaverse can restrict the range of social interactions and degrade the overall user experience, as it limits the ability to represent individuals diversely and safely (Al-kfairy et al., 2024; Bradley & Muñoz, 2024; Hennig-Thurau et al., 2022; Qin et al., 2025). Furthermore, from a policy and governance perspective, overlooking identity nuances or novel characteristics in the metaverse risks reinforcing existing hegemonic norms and anthropocentric biases, thereby perpetuating real-world biases into the metaverse. This not only impacts technology design and its intended users, but also influences the societal norms and values that these technologies reinforce and propagate (Al-kfairy et al., 2024).

*Metaverse Identity*, akin to the broader concept of *Identity*, is a multi-dimensional construct that has been examined from various disciplines and perspectives in the literature. It is predominantly considered a foundational term that underpins discussions across diverse topics in specific disciplines, such as *philosophy* (Barbara & Haahr, 2022; Van der Merwe, 2021; Mitrushchenkova, 2023; Spence, 2008), *psychology* (Barbara & Haahr, 2022; Kim, Lee, & Chung, 2023; Yang, Kim, et al., 2024), *technology* (Falchuk et al., 2018; Tao et al., 2023; Wang et al., 2024; Yang et al., 2023; Zichichi et al., 2023), *management* (Banaeian Far & Hosseini Bamakan, 2023; Bao, Nakazato, Muhammad, Javanmardi, & Tsukada, 2024; Truong et al., 2023; Wang & Wang, 2023), and law (Cheong, 2022; Mitrushchenkova, 2023; Qin et al., 2025; Wu & Zhang, 2023), which address issues and propose targeted solutions. Current research on *Metaverse Identity* can also be analyzed from multiple perspectives. The *representation perspective* treats it as the user's avatar in the metaverse, focusing on visual representation and interaction (Lee et al., 2021; Lee, Chang et al., 2023; Lee, Yang, & Hui, 2023; Park & Kim, 2022b; Saker & Frith, 2022; Schultze, 2010; Schultze & Leahy, 2009; Schultze & Orlikowski, 2010; Spence, 2008; Šímová et al., 2023). The *data perspective* considers it as the user's digital footprint and data aggregation, emphasizing data properties and the extraction of value, including specific identifiers, credentials, and biometric elements (Falchuk et al., 2018; Tao et al., 2023; Yang et al., 2023; Zichichi et al., 2023). The *social perspective* views it through the lens of the user's social roles and relationships within the metaverse, highlighting social attributes and their interactive effects (Calongne et al., 2013; Hollensen et al., 2022; Schultze, 2014, 2016; Šímová et al., 2023; Yang, Kim, et al., 2024). The *economic perspective* defines it as the user's economic entity and value carrier, often linked to blockchain technology, with a focus on asset properties and transactional potential (Lee, Yang, & Hui, 2023; Momtaz, 2022; Tao et al., 2023; Truong et al., 2023; Wang et al., 2024; Yang et al., 2022). Although these perspectives are not entirely independent and exhibit some overlap, they collectively address the primary facets of the discourse. Some researchers adopt an integrative approach, viewing *Metaverse Identity* as an extension of real-world identity into the virtual realm, thus emphasizing its multifaceted nature (Barbara & Haahr, 2022; Saker & Frith, 2022; Schultze, 2014, 2016; Won & Davis, 2023). This array of perspectives, while not exhaustive, forms the core of the scholarly discourse on this subject.

Within these contexts, *Metaverse Identity* can be defined as the multi-dimensional construct of a user's digital self, integrating personal representative attributes, data footprints, social roles, and economic elements within the metaverse. This construct encompasses various facets of an individual's presence and activities within the metaverse, encapsulating the interplay between their online persona and their real-world identity.

## 3. Core principles of metaverse identity

This section begins with an examination of recent representative cases that exemplify the manifestations and implications of *Metaverse Identity*. Building upon the previously discussed theoretical foundations and upcoming representative cases, we formulate two guiding principles essential for understanding the intrinsic characteristics, broader impacts, and challenges associated with *Metaverse Identity*. The first principle, *Equivalence and Alignment*, asserts that the analysis and understanding of *Metaverse Identity* should adhere to behavioral norms and ethical standards that mirror, or are identical to, those in the physical world. This principle is critical for recognizing the role of *Metaverse Identity* in emotional experiences, cognitive perceptions, and economic interests (Bellini, 2024; Dwivedi et al., 2022; Lake, 2019; Lee, Yang, & Hui, 2023; Mitrushchenkova, 2023). It facilitates a nuanced understanding of how identities are constructed, expressed, and perceived within the metaverse while serving as a foundation for establishing norms and safeguarding the rights inherent to these identities. The second principle, *Fusion and Expansiveness*, highlights the dynamic and evolving nature of identity within the metaverse, shaped by continuous interactions and mutual evolution. It underscores the unique role of *Metaverse Identity* in self-perception and the pursuit of personal and societal values (Al-kfairy et al., 2024; Ball, 2020; Hatada et al., 2024; Mitrushchenkova, 2023; Schultze, 2014, 2016; Yang, Kim, et al., 2024). This principle emphasizes the metaverse's capacity to enable authentic identity expressions and foster social inclusiveness. It encourages stakeholders to develop adaptable and responsive frameworks that protect users' identity rights while promoting diverse expressions and creative innovations, ensuring the metaverse's growth without the imposition of overly restrictive regulations. Together, these two principles form a robust foundation for understanding and governing *Metaverse Identity*. They address critical challenges such as identity interoperability, legal and regulatory issues, privacy management, the proliferation of deepfakes and synthetic identities, and the risks of identity fragmentation and psychological well-being, as discussed in Section 4. These principles are essential for fostering accountability and inclusivity while enabling the metaverse to evolve as a transformative socio-technical ecosystem.

*3.1. First principle - equivalence and alignment*

In recent years, the incidence of sexual assaults within the metaverse has become a pressing concern, presenting complex social and legal dilemmas (Bellini, 2024; Lee, Yang, & Hui, 2023; Schulenberg et al., 2023). For instance, in December 2021, a female metaverse user reported that three male avatars had 'touched her [her avatar] inappropriately' (Clayton, 2022). Another disturbing incident occurred in April 2022, when a female user from Japan shared via Twitter her experience of being attacked and virtually raped in VRChat while she was asleep (Anonymous, 2022). Additionally, the *Daily Mail* in the UK highlighted a particularly alarming case in January 2024 with its headline "British police probe Virtual Rape in metaverse: Young girl's digital persona 'is sexually attacked by gang of adult men in immersive video game' - sparking the first investigation of its kind and questions about extent current laws apply in online world" (Camber, 2024). The incident involved a British girl under the age of 16 whose avatar was raped by multiple avatars controlled by strangers during a virtual-reality game, raising significant legal inquiries regarding the applicability of existing laws in the metaverse.

At first glance, these cases may seem to involve merely disputes over unethical conduct and criminal actions, along with attendant legal applicability issues. However, a deeper analysis reveals that they fundamentally reflect profound challenges associated with identity recognition in the metaverse (Cheong, 2022; Kasiyanto et al., 2022; Lee, Yang, & Hui, 2023; Wu & Zhang, 2023). The reason these incidents of sexual assault attract widespread social attention and provoke a strong reaction lies in the deep psychological identification that individuals have with their avatars within the immersive metaverse platform (Lee, Yang, & Hui, 2023). If avatars were merely considered gaming tools or entertainment elements, violations against them would not significantly impact users in reality. Nevertheless, the victims' reports of discomfort and psychological trauma suggest a profound association between their avatar and their identities. This strong identification means that harm inflicted on an avatar is perceived as harm to the self, eliciting genuine emotional and psychological responses (Lee, Yang, & Hui, 2023). The legal disputes triggered by these cases also underscore the unique nature of metaverse identities in both legal and social contexts. While violations of personal rights in the physical world inevitably entail legal repercussions, in the metaverse, the identity attributes of avatars remain undefined, leading to ambiguities in rights protection (Bellini, 2024; Lee, Yang, & Hui, 2023). This ambiguity necessitates a redefinition of the identities represented by avatars. Such incidents of 'sexual assault' not only reveal the high degree of integration between avatars and users' self-presence – a psychological state in which people equate their virtual selves with their actual selves (Lee, Yang, & Hui, 2023; Schultze, 2010; Schultze & Leahy, 2009) – but also demonstrate the significant influence of metaverse identities on real-life identities, highlighting the close connection between the metaverse and real life. It is critical to acknowledge that the identity dilemmas triggered by these 'sexual assault' cases represent just one of the numerous challenges posed by entities in the metaverse. OpenAI's recent release of ChatGPT-4o,[2] an artificial intelligence chatbot capable of human-like interaction, is likely to introduce novel and complex dimensions of challenges for identity recognition in the metaverse (Hsu, 2024).

---

[2] https://openai.com/index/hello-gpt-4o/

Drawing from our theoretical analysis of *Metaverse Identity* and the challenges illustrated by cases such sexual harassment, we propose the first core principle for understanding and conceptualizing the characteristic of *Metaverse Identity*: *Equivalence and Alignment*. This principle emerges from a comprehensive consideration of multiple perspectives on *Metaverse Identity*, as discussed in the theoretical foundations, including representation, data, social, and economic aspects, addressing the identity challenges posed by metaverse entities. *Equivalence* emphasizes the parity between metaverse identities and real-world identities in aspects such as behavioral patterns, emotional experiences, cognitive perceptions, and economic interests (Lee, Yang, & Hui, 2023). *Alignment* refers to the consistency between the two in aspects concerning behavioral norms, social guidelines, and ethical standards, facilitating responsibility attribution and judgment. Both dimensions reflect our multidimensional understanding of *Metaverse Identity*. By proposing this principle, we aim to simplify the complex attributes of *Metaverse Identity*, establishing it as a foundational guideline to foster consensus and development. For instance, in addressing sexual harassment cases in the metaverse, the *Equivalence* component underscores that the psychological harm caused by virtual actions should be taken as equally significant as harm experienced in the physical world. Meanwhile, the *Alignment* component requires that real-world legal and moral standards be applied to evaluate and mitigate such behaviors. This principle underscores the unique status of *Metaverse Identity* in contemporary legal and social contexts, distinguishing the metaverse era from earlier eras such as the *Internet* and *virtual worlds*. It highlights the necessity of refining real-world regulations to address the specific challenges posed by the metaverse. Furthermore, it offers principled recommendations for policymakers, technology developers, and platform operators. Specifically, it advocates prioritizing user identity protection while promoting immersion and authenticity in the metaverse, thereby guiding necessary regulation and ethical oversight at the intersection of technology and ethics.

*3.2. Second principle - fusion and expansiveness*

Since its launch in June 2021, the *Avatar Robot Cafe* project has garnered widespread and sustained social attention due to its innovative approach and significant societal impact (Ishikawa, 2022; Kaori, 2022; Obuno, 2021; Semba, 2024; The University of Tokyo, 2023). The project's distinguishing characteristic is its innovative employment of individuals with severe physical disabilities, particularly those afflicted with neurodegenerative conditions such as Amyotrophic Lateral Sclerosis (ALS) and Spinal Muscular Atrophy (SMA), in the role of "pilots". Despite profound mobility limitations, these participants with intractable conditions utilize advanced teleoperation interfaces to remotely control robotic avatars from their residences or healthcare facilities. Initially, the project used these robotic proxies to provide customer service in cafés. Recent advancements have expanded the paradigm to encompass the remote manipulation of digital avatars, further broadening the scope of participant engagement and representation, and this novel model has attracted significant research interest (Barbareschi, Kawaguchi, Kato, Nagahiro et al., 2023; Barbareschi, Kawaguchi, Kato, Takeuchi et al., 2023; Hatada et al., 2024; Kamino & Sabanovic, 2023; Yamazaki et al., 2022). The latest publication has focused on exploring how individuals with disabilities employ robots, avatars, and a hybrid cyber–physical environment to redefine their identities (Hatada et al., 2024). This work provides crucial insights for our analysis of *Metaverse Identity*, offering valuable perspectives on the intersection of technology and identity in virtual environments.

In this study, seven participants with disabilities provided remote customer service at Avatar Robot Cafe using a combination of robots and personalized avatars. The results demonstrated that the avatars enabled participants to shift their identities during and after customer interactions. The study utilized longitudinal semi-structured interviews as one method to document these identity transformations. In this work, participant P2, diagnosed with somatoform disorder, contrasted her experiences when represented by a robot and an alpaca avatar. While using the robot, P2 felt obligated to maintain professional behavior, adhering to the belief that "failure is not an option". Conversely, when represented by the alpaca avatar, she experienced greater freedom, allowing her to adopt more playful behavior and consequently becoming a favorite among customers. P2 noted, "With the alpaca, I can tell myself, 'It cannot be helped because I am just an alpaca.' I cannot do that with OriHime (the robot)". She added, "When a customer says to my avatar, 'P2, go for it!' I enjoy it because it makes me feel like I am accomplishing what I should do as an alpaca. Just walking on two legs makes me feel proud". P2 relished interacting with others as an alpaca and greatly enjoyed her identity transformation. Participant P3, who has cerebral palsy and uses a wheelchair, identifies himself as biologically female but has a fluid gender identity (male or X). P3 created an anime-style character perceived as male or gender-neutral. This aligned with his gender preferences and pronouns. Typically subjected to perceptions of being "small and cute" due to his high-pitched voice and physical appearance, P3 noted significant shifts in how others treated him when using a taller, masculine avatar. This shift in perception facilitated a transformation in his self-expression, moving from the neutral Japanese first-person pronoun "watashi" to the more masculine informal "ore", and adopting a less direct but more assertive communication style. P3 described this as "finally becoming a boy", highlighting the profound impact of the avatar on his identity perception (Hatada et al., 2024).

The *Avatar Robot Cafe* case serves as a paradigmatic example, providing a unique perspective on another intricate dimension of metaverse identities. In this representative case, avatars act as powerful tools that enable individuals to reshape their self-identity, transcend physical limitations, and engage more fully in social activities. Avatars facilitate novel identity experiences and introduce new modes of social interaction. As the metaverse continues to evolve, we anticipate the emergence of more such scenarios, allowing a broader and more diverse group of individuals to experience these transformative interactions. This case, and others like it, leads us to propose the second core principle: *Fusion and Expansiveness*. *Fusion* refers to the profound integration of virtual and real-world identities, forming an intertwined and mutually constructive holistic self in the metaverse era. This component is important because it addresses the co-evolution of identities across virtual and real domains, allowing individuals to reconcile and integrate different facts of their existence. Within the metaverse, identity expresses needs, desires, and values that may be constrained by the physical world, serving as an extension and expansion of real-world identity. Conversely, real-world identity provides foundational context for the metaverse identity, with both complementing and enriching each other in a reciprocal relationship. *Expansiveness* underscores the unprecedented possibilities for diverse expression and the infinite extension of identity in the metaverse. This component is critical because it highlights the metaverse's ability to empower individuals to transcend their inherent traits and conditions, crafting rich, varied, and fluid alternative identities. These identities enable individuals to present themselves in different forms and fulfill diverse needs, such as social interaction and self-realization.

Schultze's theoretical research on Second Life users' identities through their avatars provides a robust foundation for the *Fusion and Expansiveness* (Schultze, 2010, 2014, 2016; Schultze & Leahy, 2009; Schultze & Orlikowski, 2010). Her theories of '*Performative Identity*' and the '*Performing Cyborgian Identity*' explicate how users express and experience identity through avatars, illustrating how the intertwining of virtual and physical identities creates a holistic, mutually constructive self (Schultze, 2014, 2016). These theories closely align with this principle, demonstrating that virtual identities not only extend and complement physical-world identities but also offer novel pathways for self-expression. In particular, the *Fusion* component highlights how virtual identities enable individuals to transcend corporeal and social constraints. The *Avatar Robot Cafe* case studies further exemplify the impact, where participants P2 and P3 manifested alternative selves through an alpaca and a masculinized anime character, respectively, thereby overcoming physical identity limitations and accessing equitable participation opportunities (Hatada et al., 2024). Schultze's emphasis on identity dynamism, which evolves through virtual-physical world interactions (Schultze, 2010; Schultze & Orlikowski, 2010), also aligns strongly with the *Expansiveness* component. Facilitated by advanced metaverse technologies, the *Fusion and Expansiveness* principle underscores unprecedented possibilities for identity expansion, allowing users to explore multifaceted identities and roles beyond physical limitations. This principle further elucidates the complexity and significance of *Metaverse Identity* construction. Metaverse identities deeply integrate with physical world identities while accentuating the metaverse's emancipatory potential from corporeal constraints. Thus, they reinforce user agency in the digital era. Empowered by technology, the malleability, fluidity, and openness of metaverse identities not only offer extensive possibilities for multidimensional self-realization but also harbor transformative potential for promoting social equity and fostering a diverse, inclusive digital ecosystem.

## 3.3. Integrating principles for metaverse identity

It is noteworthy to recognize that the principles of *Equivalence and Alignment* and *Fusion and Expansiveness* are not opposing ends of a continuum or two dimensions of an analytical quadrant. Instead, they serve as foundational pillars for the understanding, development, and governance of metaverse identities, grounded in our theoretical perspectives and representative case analyses. *Equivalence and Alignment* emphasizes the congruence and parity between metaverse identities and their physical world counterparts, particularly in terms of behavioral norms and rights protection. This principle provides a normative foundation to ensure that metaverse identities do not become spaces for irresponsibility or rights violations. It establishes an essential ethical and legal framework for integrating metaverse and physical world identities. For instance, in a metaverse educational platform, this principle ensures that students adhere to real-world academic standards. Their behavior in the virtual environment must align with established norms of academic integrity, promoting accountability in assignments, group discussions, and examinations, just as in a physical classroom. Similarly, teachers maintain their real-world authority and responsibilities, ensuring that their metaverse identities as educators comply with institutional guidelines. At the same time, the legal rights of both teachers and students must be safeguarded, including protections for privacy and safeguards against issues such as cyberbullying and sexual harassment.

Conversely, *Fusion and Expansiveness* reveals the dynamic process of mutual shaping and co-evolution between virtual and real identities through interaction. It underscores the unique role of metaverse identities in self-conception and value pursuit, as shown in Schultze's research on user identities in Second Life (Schultze, 2014, 2016). This principle highlights the agency and creativity of virtual identities within the metaverse, catalyzing individual development and social inclusivity. In the context of metaverse education, this principle manifests through the flexibility by providing to both students and teachers to explore and express their identities. For instance, students can select personalized avatars that reflect their self-identities, allowing them to engage more authentically with the learning process. Students who feel restricted or self-conscious due to physical limitations may express a more confident and engaged version of themselves through their digital personas. Similarly, teachers can adopt creative digital personas to enhance their pedagogical methods. For example, they might select avatars representing historical figures or cultural icons, enriching the learning experience and making it more immersive.

In relation to both principles, early representative features can already be observed in cases such as *HKUST Metaverse Classroom and AI Lecture* (HKUST, 2024; Pang et al., 2024). Consequently, the construction and governance of metaverse identities must prioritize *Equivalence and Alignment* as a foundational baseline to regulate and guide virtual behaviors, while simultaneously leveraging the potential of *Fusion and Expansiveness* to provide ample opportunities for identity exploration and value creation. This dual approach necessitates establishing clear boundaries and behavioral guidelines for metaverse and real-world identities when formulating policies and regulations. At the same time, it is essential to preserve sufficient flexibility and openness to encourage the expression and integration of diverse identities, fostering social inclusivity and innovation. Achieving a dynamic equilibrium between these principles is key to enabling a synergistic interaction between digital and physical existences. Such a balance ensures the orderly operation of the metaverse ecosystem while unlocking the positive potential of technological advancement.

We acknowledge that these two principles operate from a broad, macro-level perspective and retain a level of abstraction. This abstraction presents a challenge in developing conceptual frameworks for emerging fields such as the metaverse. However, as metaverse technologies continue to evolve and new application scenarios emerge, we anticipate that further empirical evidence will validate and refine these principles. For instance, empirical studies such as that of Yang, Kim, et al. (2024) provide substantive support for the *Fusion and Expansiveness* principle. Their research on VRChat users demonstrates that self-expansion experiences in metaverse environments can enhance self-esteem and life satisfaction. At the same time, their findings highlight potential risks, as identity expansion may lead to discrepancies between virtual and physical identities, which could negatively impact psychological well-being. As the field matures, we anticipate an increasing body of case studies and empirical evidence that will clarify further how these principles address the unique challenges and opportunities presented by metaverse identities. Such research will bridge the gap between conceptual frameworks and real-world applications, enabling these principles to more effectively guide the development, flourishing, and governance of metaverse identities.

## 4. Critical challenges of metaverse identity

Building upon the theoretical foundation and emblematic case analyses, we have identified two core principles – *Equivalence and Alignment* and *Fusion and Expansiveness* – to provide principled guidance necessary for effectively understanding the intrinsic characteristics and broader impacts of metaverse identities. This section will address the critical challenges that demand immediate attention, supported by these two principles. The elucidation of these five challenges serves to highlight key issues within the metaverse identity systems, thereby encouraging further interdisciplinary research and exploration.

### 4.1. Identity interoperability challenges

Interoperability challenges remain a critical obstacle to the development of the metaverse, as highlighted by both academia and industry in recent years (Ball, 2022; Wang et al., 2022; WEF, 2023a). In the systematic literature review on metaverse interoperability, Yang, Ni et al. (2024) emphasize that identity is the cornerstone of an interconnected metaverse. Without consistent and seamless identity transfer between the physical world and the metaverse, as well as across various platforms within the metaverse, true interconnectivity cannot be achieved. The principle of *Equivalence and Alignment* underscores the need for consistency and parity between metaverse identities and real-world identities, particularly in terms of behavioral norms and rights protections. Just as physical-world identities transition, authenticate, and integrate seamlessly across different contexts, metaverse identities should exhibit similar fluidity. This consistency is a crucial foundation for the metaverse to serve as a true replica of the physical world (Wang et al., 2022). Simultaneously, the principle of *Fusion and Expansiveness* highlight the dynamic symbiosis and adaptability of metaverse identities. This principle supports the creative application and diverse expression of identities across platforms, providing innovative momentum for the metaverse as a healthy extension of the physical world. Together, these two principles ensure both the standardization and coherence of metaverse identities while unlocking their creative potential, thus driving the sustainable development of the metaverse ecosystem.

The current interoperable state of *Metaverse Identity* is that most metaverse platforms continue practices from the Internet era, requiring users to create separate avatars for different platforms (Jiang et al., 2023). Identity-related data, including personal attributes, social relationships, and behavioral records, is generated by users across various platforms but is typically stored and managed in isolated silos (Wang et al., 2022). This prevents effective cross-domain data management and sharing. Existing identity systems remain predominantly platform-centric, lacking unified legal and regulatory frameworks. Definitions of rights and obligations associated with identities are ambiguous (Lee, Yang, & Hui, 2023; Yang, Ni et al., 2024). Centralized methods dominate the landscape, while decentralized identity (DID) technologies are still in their infancy and face significant technical and practical limitations (Alizadeh et al., 2022). This results in limited interoperability due to varying DID solutions across platforms, and the coexistence of centralized and decentralized methods complicates interconnected identification (Yang, Ni et al., 2024), necessitating further resolution. Additionally, the absence of unified trust mechanisms across platforms exacerbates the complexity of identity migration and association, reducing users' ability to maintain social value across multiple platforms. Economic incentives also play a critical role in restricting interoperability. Many platforms adopt "walled garden" strategies to protect their data and economic interests, restricting data flow and interoperability (Kerber & Schweitzer, 2017). These practices not only hinder the development of an open and interconnected ecosystem but also reduce industry-wide incentives for data sharing and collaboration, further fragmenting the metaverse identity environment.

Although the nascent stage of metaverse development is a primary factor contributing to these issues, it is imperative to consider the risks associated with path dependency in advance. The principle of *Equivalence and Alignment* provides a normative foundation for interoperability, ensuring consistency and security of user identities across platforms and scenarios, thereby safeguarding fundamental user experiences and rights. In contrast, the principle of *Fusion and Expansiveness* fosters innovation by supporting the diversity and dynamic evolution of interoperable identity systems, enabling creative applications and the generation of new value across platforms. Together, these two principles strike a balance between the standardization and flexibility of metaverse identities, offering a strategic framework for continuous innovation and the sustainable development of interoperable identity systems.

Coordinated efforts across technological, institutional, and industrial dimensions are essential for foster metaverse identity interoperability. Technologically, it is imperative to expedite the establishment of comprehensive norms and standards that ensure identity interoperability within the metaverse. As INTERPOL (2024) advocates, the implementation of a unique digital identity for each individual across multiple metaverses is critical. The norms and standards should encompass the creation of identity, the related data storage, interfaces and authentication protocols. Such measures are vital for establishing a standardized technical framework that supports seamless identity interoperability. Institutionally, legal and regulatory frameworks must be developed to clarify the rights and obligations of various representative identities across different contexts. These frameworks should balance interoperability with privacy and security while enabling compliant data flows. Furthermore, enhancing governance mechanisms by mobilizing platform enterprises, industry organizations, and regulatory bodies is necessary (Yang, Ni et al., 2024). This can be achieved through the establishment of a multi-stakeholder identity trust alliance, fostering robust collaboration to advance interoperability. Industrially, creating a healthy ecosystem for metaverse identity interoperability is crucial. This involves, but is not limited to, establishing data-sharing incentive mechanisms, implementing interoperability-focused privacy certifications, and promoting public education to enhance user awareness. While tackling these challenges, the principle of *Equivalence and Alignment* ensures consistency and security, and *Fusion and Expansiveness* drives flexibility and innovation. Together, these principles enable metaverse identity systems to balance interoperability with creativity, unlocking new value and enriching user experiences.

*4.2. Legal challenges of identity in metaverse*

As the metaverse continues to evolve and its applications expand, the challenges to legal boundaries are becoming increasingly apparent. In the previously discussed cases of sexual assault, victims experienced substantial psychological harassment. However, existing legal frameworks struggle to enforce traditional accountability and sanctions due to the absence of physical contact (Bellini, 2024; Cheong, 2022; Lee, Yang, & Hui, 2023). The report, "Sexual Violence and Harassment in the Metaverse: A New Manifestation of Gender-Based Harms", underscores the urgent need for updated national and international law enforcement to address this distinct form of metaverse-facilitated sexual violence and harassment (Rigotti & Malgieri, 2024). With avatar-based interactions becoming commonplace in the metaverse, the frequency of such rights conflicts and illegal acts is expected to rise. Actions that would typically incur legal penalties in the physical world – ranging from civil violations like negligence and harassment to criminal offenses such as assault and theft – pose significant challenges within the virtual settings of the metaverse. The applicability and effectiveness of existing legal frameworks in these virtual contexts remain subjects of ongoing debate (Cheong, 2022; Lee, Yang, & Hui, 2023).

In this context, applying the principle of *Equivalence and Alignment* and paralleling real-world legal structures provides a feasible approach. Scholars such as Kasiyanto et al. (2022) argue that cybercrimes within the metaverse – such as stalking, assault, child exploitation, kidnapping, intellectual property violations, and financial fraud – should be subject to legal repercussions analogous to those enforced in the physical world. They advocate for the adoption of established legal procedures from the physical domain by metaverse law enforcement agencies to effectively address these offenses. Additionally, some scholars propose extending regulatory frameworks designed for corporate entities to digital entities within the metaverse, thereby applying well-established corporate governance principles to create a suitable regulatory environment (Cheong, 2022; Day, 2009; Lake, 2019; Osborne, 2021).

The continued advancement of artificial intelligence technologies, exemplified by models such as ChatGPT-4o, is giving rise to increasingly complex digital entities, thereby posing significant challenges to existing legal frameworks. In September 2023, a court in South Korea sentenced an individual to two and a half years of imprisonment for using AI technology to create highly realistic virtual child pornography (Bae & Yeung, 2023). This landmark ruling, the first of its kind in South Korea, has garnered widespread international attention. The prosecution contended that the depiction of virtual characters, despite their fictitious nature, constituted a criminal act due to their lifelike portrayal of children (Bae & Yeung, 2023). This case has blurred the boundaries between real and virtual entities, illustrating how actions involving digital entities can carry consequences comparable to those in the physical world.

Digital entities are reshaping the concept of identity, expanding it beyond physical individuals. These AI-powered entities are capable of simulating human behavior and exhibit varying levels of interaction, autonomy, and behavioral complexity within the metaverse. They can act as virtual assistants, companions, or social media influencers (Arsenyan & Mirowska, 2021). For example, the Hong Kong University of Science and Technology has introduced AI-based digital personas, such as a simulation of Albert Einstein, to serve as virtual lecturers in its metaverse classroom (HKUST, 2024; Pang et al., 2024). Guided by the principle of *Fusion and Expansiveness*, these digital entities transcend being mere lines of code or graphical constructs. They interact with individuals and organizations, influence human behavior, and, in some cases, even represent real-world entities.

Existing legal frameworks face significant challenges in addressing issues of identity recognition, behavior evaluation, and accountability when these entities engage in harmful activities (Raposo, 2024). One proposed solution from the academic realm is to grant them the status of legal personhood, thereby holding them accountable for their actions within the metaverse (Kurki, 2019). This approach would involve integrating rights and responsibilities for these entities into existing legal framework and granting them the capacity to be a party to legal proceedings, which introduces considerable complexity. Key challenges include defining legal standards for the recognition of digital entities and addressing the impact of their actions on their controllers and other stakeholders, necessitating innovative legal and policy responses (Cheong, 2022). Aligned with the principle of *Fusion and Expansiveness*, this proposal requires a comprehensive investigation into the dynamic interplay between digital entities and real-world identities. By emphasizing the reciprocal influence and diverse representations of virtual and physical identities, this principle offers the adaptability needed to address the legal standing of digital entities. Moreover, it fosters the development of an inclusive and dynamically balanced legal framework that prioritizes both innovation and transparency.

As metaverse technologies continue to advance and the convergence of virtual and real worlds becomes more pronounced, the integration of individual identity elements into virtual spaces will intensify, thereby exacerbating legal challenges. This evolving scenario demands ongoing exploration and refinement of legal strategies (Cheong, 2022; Raposo, 2024). Addressing these emerging issues requires balancing the normative principle of *Equivalence and Alignment* with the flexible principle of *Fusion and Expansiveness*. Striking this balance facilitates the collaborative interaction between digital and physical identities, ensuring the orderly functioning of the metaverse ecosystem while harnessing the positive potential of technological advancements. Moreover, this approach fosters the creation of a diverse and inclusive digital ecosystem.

### 4.3. Privacy and identity management challenges

The metaverse presents unprecedented challenges to user privacy. A study by Nair, Guo et al. (2023), involving over 50,000 players of the popular VR Game "Beat Saber", highlights these concerns. Analysis of 2.5 million VR motion datasets through machine learning algorithms reveals that merely 100 seconds of data can uniquely identify a user with over 94.33% accuracy (Nair, Guo et al., 2023). Further research reveals that this motion data can accurately infer a wide range of personal characteristics, including biometric information such as height and wingspan, as well as demographic details like age and gender. Additionally, it can predict the user's country of origin and even clothing type. Of particular concern is the ability to accurately detect the presence of mental and physical disabilities (Nair, Rack et al., 2023). This level of precision becomes even more alarming when motion data are combined with other tracked data within the metaverse, suggesting that maintaining anonymity in such environments may be virtually impossible (Rosenberg, 2023).

Unlike traditional privacy measures that do not require sharing sensitive data, such as fingerprints, the metaverse inherently relies on the dissemination of motion data—an essential component of real-time interaction shared with all participants. Moreover, unique motion patterns analyzed across various contexts, including professional environments, social interactions, and private settings, enable the straightforward identification of individuals. The connection between physical movements and motion data further allows machine learning algorithms to correlate VR actions with real-world surveillance footage, potentially enabling real-world tracking and identification (Nair et al., 2024). In conclusion, without the development of innovative protective measures, maintaining privacy in the metaverse may prove to be an insurmountable challenge, with significant and potentially transformative implications for identity management.

Guided by the principle of *Equivalence and Alignment*, privacy protections in the metaverse should mirror the privacy rights standards established in the physical world. For example, motion data in the metaverse, which facilitates identity recognition, should be regarded as equivalent to sensitive personal information in the real world, ensuring the continuity and traceability of user rights across virtual environments. Moreover, privacy violations in the metaverse – such as data misuse or unauthorized tracking – should be subject to legal consequences comparable to those in the physical world, thereby preventing the metaverse from becoming a gray zone for accountability.

In the field of privacy protection, targeted technologies have been developed to enhance user privacy. However, these solutions currently face significant limitations, such as compromising user experience or lacking sufficient technological maturity (Nair et al., 2024). By contrast, a privacy protection framework guided by the principle of *Fusion and Expansiveness* can simultaneously safeguard user data from misuse while supporting personalized expression and diverse identity exploration. This framework, emphasizing flexibility and openness, also minimizes excessive interventions that could stifle innovation within the metaverse. For example, privacy governance in the metaverse could adopt flexible, scenario-specific protection measures. In public virtual spaces, such as virtual offices, stricter regulatory mechanisms should be implemented to prevent the misuse of behavioral data. Conversely, in private virtual spaces, such as virtual homes or rooms, enhanced privacy protections are necessary to allow users greater freedom to explore and express personalized identities (Wu et al., 2023). This scenario-based approach not only aligns with real-world privacy expectations – consistent with the principle of Equivalence and Alignment – but also facilitates multidimensional identity exploration and value creation in virtual environments, as envisioned by Fusion and Expansiveness.

By balancing the principles of *Equivalence and Alignment* and *Fusion and Expansiveness*, a dynamic and adaptive privacy governance framework can be established. On one hand, this framework extends real-world norms of behavior and rights protection into virtual spaces, ensuring the credibility and legitimacy of virtual identities. On the other hand, it fosters flexible identity expression and diversity, promoting creativity and active participation among users. This balanced approach not only supports the orderly operation of the metaverse ecosystem but also unlocks the potential of technological advancements, driving innovation and sustainable growth.

### 4.4. Challenges of deepfake and synthetic identities

A key development trend in the metaverse focuses on enhancing user experiences by making them more realistic, integrated, and immersive (Gonçalves et al., 2022), illustrating the fusion dimension of the *Fusion and Expansiveness* principle. This trend is evident in several areas. For instance, the application of advanced algorithms, such as Generative Adversarial Networks (GANs), has significantly improved the realism of rendering virtual avatars and digital entities (Koh, Tan, & Nasrudin, 2024; Saxena & Cao, 2021; Teotia et al., 2024; Tewari et al., 2020). Similarly, innovative research in facial expression (Gonzalez-Franco, Steed, Hoogendyk, & Ofek, 2020), motion capture (Jung et al., 2022), and behavioral modeling (Ahuja, Ofek, Gonzalez-Franco, Holz, & Wilson, 2021) has advanced the interactivity and authenticity of avatars and digital entities (Fysh et al., 2022; Wen, Zhou, Huang, & Chen, 2021). Furthermore, substantial progress has been made in generating and rendering virtual environments (Fink, Sosa, Eisenlauer, & Ertl, 2023; Wang, Thompson, Uz-Bilgin, & Klopfer, 2021). The rapid evolution of AI, particularly in the field of generative AI, has further simplified the creation of highly realistic and indistinguishable digital content, thereby enhancing the immersive experience of the metaverse (Murala & Panda, 2023; Qayyum et al., 2024; Ramalingam et al., 2023).

However, this technological advancement also introduces significant security risks, such as the potential for creating deepfakes or synthesizing untraceable identities, which raising serious security concerns (Tariq et al., 2023; Wu et al., 2023). Deepfakes, manipulated media generated using AI and deep learning, designed to deceive viewers, have become increasingly prevalent, sparking widespread societal discussion (Mirsky & Lee, 2021). For instance, in 2023, an AI-generated image of an explosion at the Pentagon went viral, temporarily impacting the U.S. stock market (Alba, 2023). Generative AI is reshaping the creation and expression of digital media content, while the metaverse is poised to redefine content distribution, user experiences, and interactions (Alba, 2023; WEF, 2023b). The integration of AI with VR and AR technologies is expected to amplify their collective impacts within the metaverse (WEF, 2023b). These technologies enable the creation of hyper-realistic simulations, the development of digital entities and environments, and the facilitation of deeper, more emotionally engaging interactions between users and digital entities.

Consequently, this technological expansion broadens the scope of threats, such as deepfakes (Tariq et al., 2023; Wu et al., 2023). In the absence of stringent regulatory oversight within the metaverse, users might exploit personalized identity markers – such as those derived from facial features and body forms – to commit identity theft or create synthetic identities. This presents significant risks to security and privacy, potentially resulting in more sophisticated forms of deception, defamation, and other threats within the metaverse. For example, the application of deepfake technology in virtual workplaces to fabricate digital clones of supervisors or colleagues could undermine internal trust, create confusion through deceptive work directives, or facilitate the spread of misinformation and propaganda, thereby influencing critical decisions (Tariq et al., 2023). Unfortunately, existing deepfake detection methodologies are primarily designed to address content in the physical world, such as images and videos, and do not yet adequately address the unique challenges posed by the metaverse (Tariq et al., 2023). According Wu et al. (2023) argue, there is an urgent need for deepfake recognition and prevention mechanisms specifically tailored to the metaverse.

Based on the principle of *Equivalence and Alignment*, behaviors in the metaverse should adhere to the same legal standards as those in the physical world. For instance, when users employ deepfake technology to impersonate others, such actions should be treated as equivalent to identity theft or fraud in the physical world and subject to legal accountability. To address these concern, clear legal frameworks must be established to explicitly prohibit the use of virtual avatars for fraud, defamation, or the dissemination of false information. Furthermore, platforms should be mandated to implement robust identity verification technologies to ensure the authenticity and traceability of digital content. Looking ahead, addressing these challenges will require a dual approach: developing advanced technological solutions for identity verification and content authenticity, and establishing explicit regulatory frameworks that define legal boundaries for virtual avatars and digital content. These frameworks should also address critical issues such as privacy protection and misuse prevention.

### 4.5. Identity fragmentation and psychological well-being challenges

The metaverse offers unprecedented opportunities for diverse and expansive identity representation, enabling individuals to transcend the limitations of the physical world. While this diversity in representation can foster creativity and inclusivity— exemplified by the case of the *Avatar Robot Cafe*, where marginalized groups gained equal participatory opportunities (Hatada et al., 2024)—it also introduces significant risks, such as identity fragmentation and confusion. Recent research by Yang, Kim, et al. (2024), which involved a two-wave panel study on the VRChat platform, highlights these concerns. The findings indicate that while the expansion of identity representation can enhance self-esteem and life satisfaction, these benefits may diminish when inconsistencies in self-concept arise. A notable historical example is the 2000 incident in which a child, after prolonged engagement with a video game, experienced severe identity dissonance. Believing himself to be the game's protagonist, he ventured out at night to "fight enemies and save the princess", mimicking the game's narrative. This blurring of reality and fantasy ultimately required medical intervention to restore his sense of identity and reality (Banerji, 2023). Such incidents underscore the importance of addressing the psychological risks associated with identity exploration in virtual environments.

As the metaverse evolves, addressing its psychological impacts and implementing protective measures to safeguard identity integrity, particularly among children and vulnerable populations, becomes increasingly urgent. Identity fragmentation within the metaverse can significantly impair users' mental health and social relationships. The ease of creating and switching between diverse, unintegrated identities may reduce reliance on real-world personas, thereby exacerbating issues of identity fragmentation (Yang, Kim, et al., 2024). These "consequence-free" virtual spaces (Turkle, 1999), which often diverge sharply from reality, pose specific risks. Such environments can lead to self-misalignment, where users perceive their virtual identities as superior, causing them to prioritize virtual interactions over real-world relationships and resulting in alienation from the physical world (Banerji, 2023). This alienation is particularly concerning for children and adolescents, who may increasingly immerse themselves in the metaverse, leading to social withdrawal, depression, and antisocial behavior (Koohsari et al., 2023). Given the well-documented adverse effects of social media, video games, and mobile devices, the immersive and escapist nature of the metaverse warrants critical examination to mitigate its potentially detrimental impacts on psychological and social well-being (Banerji, 2023; Koohsari et al., 2023).

Research on multiple cultural identities indicates that the presence of multiple identity representations is not inherently problematic. Instead, the critical factor lies in how these identities are related, integrated, and managed (Yang, Kim, et al., 2024). As the metaverse continues to evolve and identity-related issues become more prominent, further research and practical interventions are essential. Future studies should prioritize developing strategies to assist users in managing their multiple identities and preventing identity fragmentation, thereby mitigating associated psychological and social challenges. Given that these concepts remain in their early stages, proactively recognizing and addressing identity fragmentation is vital for effectively tackling the emerging challenges of the metaverse.

Guided by the principle of *Equivalence and Alignment*, identity management in the metaverse should adhere to the behavioral norms, ethical standards, and legal responsibilities of the physical world. This approach minimizes conflicts between virtual and real identities while providing stronger safeguards for psychological well-being. Simultaneously, the principle of *Fusion and Expansiveness* emphasizes the dynamic integration and complementarity between virtual and real identities. By helping users achieve balance among their multiple identities, this principle enables the metaverse to maximize its inclusivity and innovative potential. Policymakers should prioritize the development of an identity management framework that balances regulatory constraints with flexible expression, while safeguarding users' mental health and social connections. Specifically, initiatives in education, technological support, and mental health interventions are essential to help users effectively manage and integrate their multiple identities, prevent the adverse effects of identity fragmentation, and harness the metaverse's potential to enhance social inclusivity and individual value realization.

## 5. Conclusion and limitation

The metaverse, an emerging realm that merges virtual and real elements, significantly influences identity expression, social interaction, and human development trajectories. At the core of the metaverse lies the construction of identity, encompassing the representation of personal traits, management of data footprint and social roles, identification of avatars and other digital entities, and economic elements. These elements are closely tied to users' sense of belonging, privacy, security, and trust. Despite its significance, scholarly research on metaverse identities remains in its early stages and lacks a comprehensive theoretical framework. This paper reviews key literature to explore the concept of *Metaverse Identity* from multidisciplinary and multi-perspective viewpoints, identifying the research gap. Unlike traditional Internet and virtual worlds settings, the metaverse offers more immersive and lifelike experiences, transforming identity from a mere theoretical construct into a tangible, experiential reality. This transformation underscores the tension between benefits – such as enhanced self-expression – and challenges like privacy protection.

Metaverse identities are characterized by their multilayered, dynamically evolving, and diverse nature. This study proposes a multidimensional framework for *Metaverse Identity*, encompassing its definition, principles, and critical challenges. Drawing on existing theoretical foundations, *Metaverse Identity* is defined as a multidimensional construct of a user's digital self, integrating personal representative attributes, data footprints, social roles, and economic elements within the metaverse. Based on this definition, this paper proposes two core principles: *Equivalence and Alignment* and *Fusion and Expansiveness*. The first principle advocates for consistency between the metaverse and real-world identities in terms of behavioral norms and social standards, which is essential for developing conduct guidelines and protecting rights. The second principle stresses the need for deep integration and extensive expansion of metaverse identities, going beyond real-world constraints to meet diverse needs and promote inclusive participation. Effective governance in the metaverse necessitates a dynamic balance between these principles, ensuring fairness while encouraging diverse expressions and innovative developments. This paper further identifies five key challenges to the development of metaverse identities: interoperability, legal boundaries, privacy and identity management issues, the risks associated with deep fakes & synthetic identities, and identity fragmentation impacting psychological health. Building on the principles outlined, this paper offers strategic recommendations to address these challenges. By constructing this multidimensional framework, this study addresses the critical theoretical gap in understanding *Metaverse Identity*. This framework not only provides a foundation for advancing systematic and empirical research but also serves as a basis for formulating policies and governance strategies to tackle the multidimensional challenges of this rapidly evolving digital domain.

This study has several significant limitations. Firstly, while the authors have clarified the rationale for selecting representative cases – emphasizing that these cases serve to concretize abstract principles rather than inductively derive them – the representativeness of the case studies remains limited. Given the metaverse's diverse and complex ecosystem, the analysis may not fully capture the wide spectrum of identity practices. Furthermore, due to the nascent stage of metaverse development, the number of available cases is constrained. Future research should broaden the scope of case studies to enhance the generalizability and relevance of the findings. Secondly, the empirical foundation of this study requires further strengthening. Research on *Metaverse Identity* is still in its infancy and lacks comprehensive data on user behavior and psychological responses. Currently supported primarily by a literature review and qualitative analysis, this study underscores the need for more rigorous empirical research. Future efforts should employ observational, survey, and experimental methodologies to explore identity construction behaviors and perceptions in the metaverse, thereby solidifying the theoretical framework. Lastly, the theoretical principles introduced – *Equivalence and Alignment*, and *Fusion and Expansiveness* – require further refinement and operationalization. While these principles offer valuable insights into understanding, developing, and regulating identity in the metaverse, they remain broad and abstract. Their practical application in governance and policy-making may face challenges. Future research should aim to refine these principles and develop detailed, actionable guidelines to inform the governance and design of metaverse identity frameworks.

Looking ahead, the metaverse, as an emerging socio-technical ecosystem in the digital age (Lee et al., 2021; Lee, Yang, & Hui, 2023; Peukert et al., 2024; Wang et al., 2022; Yang, Ni et al., 2024), is poised to significantly transform human life and developmental practices. The manner in which individuals construct and manage their identities across the physical world and the metaverse is critical for sustainable human development. Addressing these challenges require a unified effort from academia, industry, and policy sectors to promote interdisciplinary, multi-perspective research, aimed at fostering a robust and inclusive metaverse environment. Such collaborative initiatives are vital to fostering a robust and inclusive metaverse environment, providing both intellectual and practical contributions to the evolution of this new digital era.

## CRediT authorship contribution statement

**Liang Yang:** Writing – review & editing, Writing – original draft, Methodology, Formal analysis, Conceptualization. **Yan Xu:** Project administration. **Pan Hui:** Supervision.

**Acknowledgments**

The authors thank Canhui Liu from the University of Cambridge for an earlier discussion, the committee members of the 24th Biennial Conference of the International Telecommunications Society (Seoul, Korea) for awarding an earlier version of this work the First Prize Student Paper Award, and the anonymous reviewers for their constructive and detailed feedback. This research was partially supported by the sub-project *3D Virtual Scenes and Avatars*, funded by the Guangzhou Municipal Nansha District Science and Technology Bureau (Contract No. 2022ZD012); the sub-project *AI-driven Intelligent Avatars* funded by HSBC (Project No. L0562); and a fund from the Hong Kong University of Science and Technology and the Education University of Hong Kong Joint Centre for Artificial Intelligence (JCAI).

L. Yang et al.                                                                                                                                    Telecommunications Policy 49 (2025) 102906